\documentclass{aa501}
\newlength{\mylength}
\usepackage{graphicx}
\newcommand{\slf}[2]{\mbox{$\,^{#1}\!\!/\!_{#2}$}}
\newcommand{\lta}{\la}
\newcommand{\gta}{\ga}
\newcommand{\rd}{{\rm d}}
\newcommand{\calB}{{\cal B}}
\newcommand{\calR}{{\cal R}}
\newcommand{\mue}{\mu_{_{\rm{}E}}}

\begin{document}

\title{Orbital decay of satellites crossing an accretion disc}


\author{V. Karas \and  L. \v{S}ubr}


\institute{Astronomical Institute, Charles University Prague,
 V~Hole\v{s}ovi\v{c}k\'ach~2, CZ-180\,00~Praha, Czech Republic\\
 \email{vladimir.karas@mff.cuni.cz; subr@aglaja.ms.mff.cuni.cz}}

\date{Received 16 January, 2001; accepted 8 June, 2001}

\abstract{Motion of stellar-mass satellites is studied around a massive
compact body which is surrounded by a gaseous slab of a stationary
accretion disc. The satellites suffer an orbital decay due to
hydrodynamical interaction with the disc medium (transitions across the
disc, gap opening in the disc, density waves) and gravitational
radiation. Arbitrary orbital eccentricities and inclinations are
considered, and it is observed how the competing effects depend on the
parameters of the model, namely, the mass and compactness of the
orbiters, the osculating elements of their trajectories, and surface density
of the disc. These effects have a visible impact on the satellites
long-term motion, and they can produce observational consequences 
with respect to galactic central clusters. It is shown that the
satellite-disc collisions do not impose serious restrictions on the
results of gravitational wave experiments if the disc medium is diluted
and the orbiter is compact but they are important in the case of environments
with relatively high density. We thus concentrate on application
to accretion flows in which the density is not negligible. We discuss the
expected quasi-stationary structure of the cluster that is established 
on sub-parsec scales within the sphere of gravitational influence of the 
central object. Relevant to this region, we give the power-law slopes 
defining the radial profile of modified clusters and we show that their 
values are determined by satellite interaction with the accretion flow
rather than their initial distribution.
}
\maketitle
\newcounter{mynumber}
\section{Introduction}
There is ample evidence that central regions of galaxies consist of
three main constituent parts: a supermassive dark object (presumably a
black hole), a dense cluster of stars, and an accretion flow with
disc-type geometry (for a review, see Rees 1998; Kato, Fukue \&
Mineshige 1998). Related to these three ingredients are
typical length-scales:
(i)~$r_{\rm{g}}=GM/c^2\approx1.5\times10^{13}\,M_8\,\rm{cm}$
(gravitational radius of the central black hole expressed in terms of
its mass $M_8{\equiv}M/10^8M_{\sun}\lta1$).
(ii)~$r_{\rm{c}}=GM/\sigma^2\approx0.5M_8\sigma_{1000}^2\,$pc, the
size of an inner star-cluster (where motion of its members is governed
mainly by the black-hole gravitational field) written in terms of
stellar velocity dispersion $\sigma_{1000}\equiv\sigma/(1000\,$km/s).
A typical number of
stars in this region can be of the order $N\approx10^5$. Finally,
(iii)~$r_{\rm{d}}\approx10^4 r_{\rm{}g}$ (the size of the disc). For the
purpose of this paper these stars play a role of stellar-mass satellites
orbiting the center. The disc itself will be described in the thin-disc
approximation with the total mass $M_{\rm{d}}{\ll}M$.

In this paper we assume that the three mentioned subsystems coexist and
they form an integral structure for a sufficiently long period of time.
This is a reasonable assumption for a range of model parameters
(discussed below), and the resulting geometry captures interesting
effects of realistic systems, although it is simplified in several
respects. The three subsystems are of very different nature, so that
their symbiosis and mutual interactions pose a complex, ecological
problem (e.g., Shlosman \& Noguchi 1993; van der Marel \& van den Bosch
1998; Zwart, Hut \& Verbunt 1997). We confine ourselves to a simple
analysis in which the gravitational field of the system is dominated by
the central compact mass while long-term dynamics of the satellites is
influenced by dissipation in the disc medium and by gravitational
radiation. The present paper is motivated mainly by indications of
substantial concentrations of both stars and interstellar medium in the
galactic central parsec. For example, a toroidal system of stars and gas
has been proposed as a plausible explanation of brightness peaks
observed in the nucleus of M31 (Tremaine 1995; Bekki 2000). However, as
we want to discuss also the regime in which inspiraling satellites are
influenced by gravitational radiation in the field of the massive
centre, we need to consider sub-parsec scales which are not yet resolved
observationally. This appears relevant for the continuation (towards small
length-scales) of the velocity dispersion versus mass relation (Ferrarese \&
Merritt 2000), which is so important in the discussion of the masses
of supermassive black holes. Finally, core-collapsed globular clusters
could be another class of relevant objects on scales of masses
$\approx10^4M_{\sun}$, i.e.\ much less than galactic cores.

We proceed along the same course as several recent works in which further
references can be found: Rauch (1999) studied stellar dynamics near a
massive black hole, including the effects of general relativity on
stellar trajectories; \v{S}ubr \& Karas (1999) examined the long-term
evolution of orbital parameters of a satellite colliding with a thin
Keplerian disc; Narayan (2000) estimated the relative importance of
hydrodynamic drag versus gravitational-radiation decay of the satellite
orbit (this author was interested primarily in the case of a body
embedded in an advection-dominated flow very near the central hole);
Collin \& Zahn (1999), on the other hand, explored the fate of
satellites moving on the periphery of the disc and at intermediate
distances ($\approx1\,$pc for $M_8\approx1$) where mutual interaction of
the disc matter with the satellites triggers the starburst phenomenon
and provides a mechanism of metal enrichment; finally, Takeuchi,
Miyama \& Lin (1996) and Ward (1997) discussed the problem of satellite
migration via gap formation and density waves in the disc, and they
gave a detailed overview of related topics which have been widely
discussed also by others.

The orbital motion of satellites forming the central cluster is
dominated by gravity of the central body; however, the long-term
evolution of their trajectories is affected also by many other effects
such as gravity of the background galaxy (e.g., Sridhar \& Touma 1999),
dynamical friction and tidal interactions (Colpi, Mayer \& Governato
1999), self-gravity of the disc matter (Shlosman \& Begelman 1989;
Vokrouhlick\'y \& Karas 1998), repetitive collisions with an accretion
disc (Syer, Clarke \& Rees 1991; Rauch 1995), and gravitational
radiation losses (Peters \& Mathews 1963). The interaction between the
members of the cluster and the disc are of interest also because they
may gradually change the composition of the disc (Artymowicz, Lin \&
Wampler 1993), speed up the evolution of the satellite itself (Collin \&
Zahn 1999), and they can enrich the environment outside the disc plane
(Zurek, Siemiginowska \& Colgate 1994; Armitage, Zurek \& Davies 1996).

Using simple estimates we compare the energy and angular
momentum losses via gravitational radiation against the hydrodynamical
drag acting on the satellites. We assume that their orbits have,
initially, arbitrary eccentricities and nonzero inclinations with
respect to the disc plane -- the situation which is complementary to the
case of planetary formation and their migration inside the disc. For the
subsequent (low-inclination) stages of the orbital evolution we employ a
simple prescription in which gravitational radiation losses still tend
to bring the orbiter towards the centre while hydrodynamical effects are
approximated in terms of motion through gas.
 
We build our discussion on a model description (\v{S}ubr \& Karas
1999) in which the satellite suffers from dissipation of the orbital
energy during repetitionary star-disc interactions but the disc itself
does not change with time. We show evolutionary tracks followed by the
satellite in the parameter space of its orbital osculating elements. The
results are found quite sensitive to actual values of the disc density,
the mass of the satellite, and to several other parameters of the model.
We discuss the dynamical evolution of a cluster of satellites which
gradually departs from its initially spherical shape and evolves into a
flattened system. The initial stage of the evolution (when most of the
orbits are still inclined with respect to the disc) is similar to the
situation discussed by Pineault \& Landry (1994) and Rauch (1995), 
but for the later stages we must adapt the prescription for
star-disc interactions. See also Ivanov, Papaloizou \& Polnarev (1999) 
for a complementary discussion of the gradual evolution of 
a circumbinary disc structure.

\section{Satellite motion influenced by the disc}
\label{sec:individual}
The satellite body is assumed to cross the disc at supersonic velocity
with a high Mach number, ${\cal{}M}\approx10^2$--$10^3$. We recall 
(Syer et al.\ 1991; Vokrouhlick\'y \& Karas 1993; Zurek et al.\ 1994) that
passages last a small fraction of the orbital period at the corresponding
radius where they occur, and they can be considered as instantaneous,
repetitive events at which the passing satellite expels from the disc
some amount of material that lies along its trajectory. In terms of
inclination angle $i$ and the disc thickness $h$, the typical ratio of the
two periods is $\delta{\approx}h/(r\sin{i})$; this quantity is assumed
to be less than unity. Hence, for a geometrically thin disc
($\epsilon\equiv{h/r}\ll1$), its influence upon inclined stellar orbits
can be treated as tiny kicks (impulsive changes of their energy and
momentum) at the points of intersection with the plane of the disc.
This can be translated using the relation for the speed of sound,
$c_{\rm{}s}=\epsilon{\Omega}r$, to the claim that the motion across the
disc is indeed supersonic, which is also required for consistency.
Smallness of $\delta$ means that the satellite remains outside the disc
for most of its revolution around the centre; naturally this condition
cannot be ensured at final stages when $i$ is very small.

\subsection{Collisions with the disc}
Repetitive collisions lead to the gradual change of the orbiter's velocity
$\vec{v}\rightarrow\vec{v'}$, which can be expressed by momentum
conservation (\v{S}ubr \& Karas 1999):
\begin{equation}
 (A+1)\,\vec{v'}=v_r\,\vec{e_r}+v_{\vartheta}\,\vec{e_\vartheta}
 +\left(v_\varphi+Av_{_{\rm{K}}}\right)\vec{e_\phi}
\label{eq:velocity_change}
\end{equation}
in spherical coordinates ($\vartheta=\pi/2$ is the disc plane). Here,
$A(r)\,\equiv\,{\Sigma_{\rd}}v_{\rm{rel}}
\Sigma_{\ast}^{-1}v_\vartheta^{-1}$, $v_{\rm{rel}}$ is the relative
speed between the orbiter and the disc matter, $\Sigma_{\ast}$ is the
column density characterizing the compactness of the orbiter and defined by
$\Sigma_{\ast}=M_\ast/\left(\pi{R^2_{\ast}}\right)$ (quantities denoted
by an asterisk refer to the orbiter, $M_\ast{\ll}M$), and
$\Sigma_{\rd}(r)$ is the disc surface density. Rotation of the disc is
assumed Keplerian, $v_\phi\,\equiv\,v_{_{\rm{}K}}=\sqrt{GM/r}$.

Let us consider a satellite on an orbit with semi-major axis $a$,
eccentricity $e$, inclination $i$, and longitude $\omega$ of the
ascending node. Eq.~(\ref{eq:velocity_change}) implies a set of
equations which can be solved numerically in terms of the orbiter's
osculating elements, while analytical solutions are possible in special
cases (\v{S}ubr \& Karas 1999). As a useful example, we assume a
power-law surface-density distribution in the form
$\Sigma_{\rd}=K\left(r/r_{\rm{}g}\right)^s\Sigma_{\sun}$
($K={\rm{const}}$) and we adopt a perpendicular orientation of the orbit,
$\cos\omega=0$. We find
\begin{equation}
a = C_1\,x_+\left(x_+^3+C_2\,x_-\right)^{-1},\quad
y = 1+C_2\,x_-\,x_+^{-3},
 \label{eq:az1}
\end{equation}
where $x_\pm\equiv1\pm{x}$, $x\equiv\cos{i}$, $y\equiv1-e^2$. Strictly
speaking, this formula concerns only the case of orbits intersecting the
disc at two points with identical radial distances from the center but
it can be used also as an approximation for orbits with arbitrary
orientation. We remark that the relative accuracy $\Delta{a}/a$ of the
determination of the semi-major axis is better than 15\% with reasonable
density profiles ($s$ of the order of unity); numerical computations are
not limited by assumptions about $\omega$ imposed in eq.~(\ref{eq:az1}).
Furthermore, one can write
\begin{equation}
 A=\frac{K\Sigma_{\sun}}{\Sigma_{\ast}}
\left(\frac{ay}{r_{\rm{g}}}\right)^s\;\sqrt{3-y-2x \over 1-x^2}\,.
\end{equation}

Integration constants $C_{1,2}$ are to be determined from initial 
values of $a=a_0$, $x=x_0$, and $y=y_0$. Then, the temporal history 
is obtained by integrating over the orbital period, 
${\rd}t=2{\pi}a^{3/2}/\sqrt{GM}$, in the form
\begin{equation}
 t=\frac{2\pi}{K\sqrt{GM}}\frac{\Sigma_{\ast}}{\Sigma_{\sun}}
 \int_{x_0}^{x} {a(\bar{x})^{3/2}
 \left[ a(\bar{x})
 y(\bar{x}) / r_{\rm{g}} \right]^{-s}\rd \bar{x} \over
 \sqrt{ \left( 3 - y(\bar{x}) -
 2\bar{x} \right) \left( 1 - \bar{x}^2 \right) }} \,.
 \label{eq:tz1} 
\end{equation}
Here, a factor missing in eq.~(19) of \v{S}ubr \& Karas (1999) is
corrected (no other equations and graphs were affected by that
omission). The orbital decay manifests itself in the gradual decrease of
$a$, $e$ and $i$, for which surface density of the disc is the main
factor. The time derivative of the semi-major axis is
\begin{equation}
 \dot{a}_{\rm{}col}={\calB}y^{-q_4}
 \left[\frac{\Sigma_{\ast}}{\Sigma_{\sun}}\right]^{-1}
 \sqrt{\frac{3-y-2x}{y(1-x^2)}} \; (2-x-y) \,,
 \label{dadtcoll}
\end{equation}
where
\begin{equation}
 \calB=-BcM_8^{q_1}\mue^{q_2}\!
 \left[\frac{\alpha}{0.1}\right]^{q_3}
 \left[\frac{a}{r_{\rm g}}\right]^{-q_4}
\end{equation}
and $\mue\equiv\dot{M}/(0.1\dot{M}_{\rm E})$ is the accretion rate in units
of Eddington accretion rate (with a 10\% efficiency factor introduced).
The factor $B$ and power-law indices $q_{1\ldots4}$ are determined by
details of the particular model adopted to quantify the disc
properties. Table~\ref{tab1} gives the values relevant for different
regions of the standard disc as well as for the gravitationally unstable
outer region (Collin \& Hur\'{e} 1999). Notice that the algebraic
functional form of radial dependencies remains identical in all these
cases (a power-law), and we can use it with convenience also later for
different prescriptions of the satellite-disc encounters.

\begin{table}[bt]
\centering
\caption{Parameters in eqs.~(\ref{dadtcoll}), (\ref{dadtgap}) and
 (\ref{dadtdw}) describing the orbital decay in the case of different
 regimes and for different disc models.
\label{tab1}}
\begin{tabular}{ll|lcccc} \hline
 Regime \rule[-1ex]{0mm}{3.5ex} & Disc 
  & \hspace*{3ex}$B$ & $q_1$ & $q_2$ & $q_3$ & $q_4$ \\ \hline
 {\sf\hspace*{2.5ex}col} \rule[0ex]{0mm}{2.5ex}&
 (i) & $2.3\times10^{-9}$ & $0$ & $-1$ & $-1$ & $-1$ \\
 {\sf\hspace*{2.5ex}col}  &
 (ii) & $2.9\times10^{-5}$ & $\slf{1}{5}$ & $\slf{3}{5}$
  & $\slf{{-4}}{5}$ & $\slf{11}{10}$ \\
 {\sf\hspace*{2.5ex}col}  &
 (iii) & $1.1\times10^{-4}$ & $\slf{1}{5}$ & $\slf{7}{10}$
  & $\slf{-4}{5}$ & $\slf{5}{4}$ \\
 {\sf\hspace*{2.5ex}col}  &
 (iv) & $4.05$ & $\slf{-9}{7}$ & $\slf{1}{7}$ & $0$ & $\slf{37}{14}$ \\
 {\sf\hspace*{2.5ex}col}  &
 (v) & $1.1\times10^{-2}$ & $-1$ &  \slf{1}{9} & $0$ & $\slf{13}{6}$ \\

 {\sf\hspace*{2.5ex}gap} \rule[0ex]{0mm}{3ex}&
 (i) & $1.3\times10^{-3}$ & $0$ & $2$ & $1$ & $\slf{5}{2}$ \\
 {\sf\hspace*{2.5ex}gap}  &
 (ii) & $7.4\times10^{-8}$ & $\slf{-1}{5}$ & $\slf{2}{5}$
  & $\slf{4}{5}$ & $\slf{2}{5}$ \\
 {\sf\hspace*{2.5ex}gap}  &
 (iii) & $2.8\times10^{-8}$ & $\slf{-1}{5}$ & $\slf{3}{10}$
  & $\slf{4}{5}$ & $\slf{1}{4}$ \\
 {\sf\hspace*{2.5ex}gap}  & (iv)   
  & $6.1\times10^{-13}$ & $\slf{9}{7}$ & $\slf{6}{7}$ & $0$ & $\slf{-8}{7}$ \\
 {\sf\hspace*{2.5ex}gap}  & (v)
  & $2.2\times10^{-10}$ & $1$ & $\slf{8}{9}$ & $0$ & $\slf{-2}{3}$ \\

 {\sf\hspace*{2.5ex}dw} \rule[0ex]{0mm}{3ex}&
 (i) & $1.8\times10^{-18}$ & $0$ & $-3$ & $-1$ & $-5$ \\
 {\sf\hspace*{2.5ex}dw} &
 (ii) & $1.3\times10^{-10}$ & $\slf{2}{5}$ & $\slf{1}{5}$
  & $\slf{-3}{5}$ & $\slf{-4}{5}$ \\
 {\sf\hspace*{2.5ex}dw} &
 (iii) & $4.8\times10^{-9}$ & $\slf{2}{5}$ & $\slf{2}{5}$
  & $\slf{-3}{5}$ & $\slf{-1}{2}$ \\
 {\sf\hspace*{2.5ex}dw} &
 (iv) & $8.2\times10^{-7}$ & $\slf{-5}{7}$ & $\slf{-1}{7}$
  & $0$ & $\slf{5}{14}$ \\
 {\sf\hspace*{2.5ex}dw} \rule[-1.5ex]{0mm}{2ex} &
 (v) & $2.9\times10^{-4}$ & $-1$ & $\slf{-1}{9}$ & $0$ & $\slf{5}{6}$ \\
\hline
\end{tabular}
\par
\begin{list}{}{}
\item[Notation used in table.]~
\item[Disc models:]~(i)~\ldots~Standard disc with $p=p_\mathrm{rad}$,
 $s=3/2$; (ii)~\ldots~Standard disc $p=p_\mathrm{gas}$, $s=-3/5$,
 electron scattering opacity; (iii)~\ldots~The same as (ii) but with
 $s=-3/4$ and free-free opacity; (iv)~\ldots~Marginally unstable
 self-gravitating disc (solar metallicity, optically thick), $s=-15/7$;
 (v)~\ldots~The same as (iv) but for zero metallicity, optically thin
 medium, $s=-5/3$; cf.\ Sec.~\ref{evolution} for further details.
\item[Regimes of orbital decay:]~Orbital decay dominated by
 star-disc collisions~({\sf{}col}), by gap formation in the disc
 ({\sf{}gap}), and  by density waves ({\sf{}dw}), respectively.
\end{list}
\end{table}

\subsection{An orbiter embedded in the disc}
As drag is exerted on the satellite body, its orbit becomes circular
and declined in the disc plane. Within this framework, orbital
eccentricity and inclination are expected to reach zero values in the
disc, so that quasi-circular trajectories are relevant near the center.

The orbit evolution is thus reduced to the situation which was addressed
by several people (e.g., King \& Done 1993; Takeuchi et al.\ 1996; Ward
1997; Ostriker 1999) in connection with formation and subsequent
migration of bodies inside the disc. Two basic modes can be
distinguished according to the disc properties and the orbiter mass.
First, a gap is cleared in the disc if the satellite's Roche radius
exceeds the disc thickness,
$r_{_{\rm{}L}}\approx(M_\ast/M)^{1/3}r{\gta}h$, and simultaneously
$M_{\ast}{\gta}M_{\rm{}gap}{\approx}\sqrt{40\alpha}\,\epsilon^{5/2}M$
(Lin \& Papaloizou 1986). Motion of the satellite is then coupled with
the disc inflow, so that
\begin{equation}
 \dot{a}_{\rm{}gap}={\calB}.
 \label{dadtgap}
\end{equation}
On the other hand, if the satellite is unable to create the gap, the
gas drag is imposed on it through quasi-spherical accretion. 
The resulting radial drift is weaker by a factor $\epsilon\ll1$
than the drift caused by density-wave excitation (Ward
1986; Artymowicz 1994). Hence, such a satellite migrates inward mainly
due to the latter effect on the time-scale
\begin{equation}
t_{\rm{}dw}= \left(CM_{\ast}\Omega\right)^{-1}\epsilon^2
 \left(\frac{M^2}{\pi{}r^2 \Sigma_{\rd}}\right),
\label{eq:time_wave}
\end{equation}
where $C$ is a dimensionless constant of the order of unity. Using
eq.~(\ref{eq:time_wave}) and ${\rd}a/{\rd}t{\approx}r/t_{\rm{}dw}$ we
obtain
\begin{equation}
\dot{a}_{\rm{}dw}=\frac{M_\ast}{M_{\sun}}\,{\calB}.
 \label{dadtdw}
\end{equation}
Substantial differences in the satellite migration are thus introduced
in the model already within this very simplified picture where the
process of satellite sinking is driven by the gas medium. For the
region of the gas pressure dominated standard disc and for a self-gravitating
zero-metalicity model (v), the chance of opening the gap increases with
decreasing radius. The situation is opposite in the case (i) (radiation
pressure dominated) and for the solar-metallicity disc~(iv).

\subsection{The orbital decay due to gravitational radiation}
\begin{figure}[t]
\centering
\includegraphics[width=\mylength]{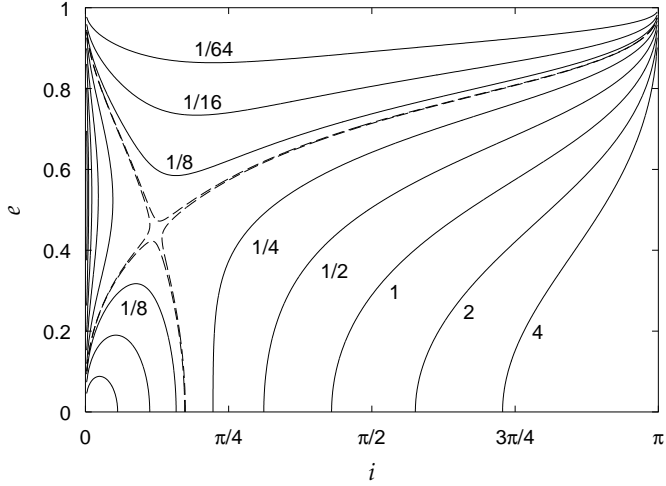}
\caption{Contours of $f$ are plotted as a function of inclination and 
 eccentricity for the case (ii). Function $f(i,e)$ characterizes the 
 relative importance of energy losses in eq.~(\protect\ref{eq:f_middle}). 
 Values of $f$ are indicated with contour lines. A saddle-type point develops
 between two neighbouring contours, $f=0.150$ and $f=0.152$, which 
 are plotted with dashed lines. Its exact location depends on the disc model
 but the overall picture remains very similar in other cases, too.}
 \label{fgr2}
\end{figure}
The orbiting companion emits continuous gravitational radiation whose
waveforms are of particular relevance for gravitational wave searches
from compact binaries in the Milky Way. Possible approaches to their
observational exploration have been discussed by several people (e.g.,
Nakamura, Oohara \& Kojima 1987; Dhurandhar \& Vecchio 2001; Hughes
2001). The average rate of energy loss which the orbiter experiences
via gravitational radiation over one revolution can be written in terms
of orbital parameters (Peters \& Mathews 1963),
\begin{equation}
 \dot{E}_{\rm{}gw}=\frac{32}{5}\frac{G^4}{c^5}
 \frac{M^3M_{\ast}^2}{a^5\left(1-e^2\right)^{7/2}}
 \left(1+\textstyle{\frac{73}{24}}e^2
 +\textstyle{\frac{37}{96}}e^4\right).
 \label{degw}
\end{equation}
Corresponding to eq.~(\ref{degw}) are the change of semi-major axis
\begin{eqnarray}
 \dot{a}_{\rm{}gw} &=& -1.28\times10^{-7}c
 M_8^{-1}\frac{M_{\ast}}{M_{\sun}}
 \left[\frac{a}{r_{\rm g}}\right]^{-3}
\nonumber \\ & & \times
 \frac{1+\textstyle{\frac{73}{24}}e^2+
 \textstyle{\frac{37}{96}}e^4}{(1-e^2)^{7/2}}\,,
\end{eqnarray}
and the loss of angular momentum 
\begin{equation}
 \dot{L}_{\rm{}gw}=
 \frac{32}{5}\frac{G^{7/2}}{c^5}
 \frac{M^{5/2}M_\ast^2}{a^{7/2}\left(1-e^2\right)^{2}}
 \,\left(1+\textstyle{\frac{7}{8}}e^2\right).
\label{dlgw}
\end{equation}
The above formulae (\ref{degw})--(\ref{dlgw}) assume that the
satellite star follows an eccentric orbit in Schwarzschild geometry of
the central massive body. Gravitational radiation losses compete with
those caused by star-disc encounters. We are thus interested in the
relative importance of these mechanisms, which is characterized by the
ratio
${\calR}_{\rm{}col/gw}=\left[{\rd}a/{\rd}t\right]_{\rm{}col}/
\left[{\rd}a/{\rd}t\right]_{\rm{}gw}$:
\begin{equation}
{\calR}_{\rm{}col/gw}=\frac{5K}{32\pi}\frac{M}{M_\ast}
 \left[\frac{\Sigma_{\ast}}{\Sigma_{\sun}}\right]^{-1}
 \left[\frac{a}{r_{\rm{g}}}\right]^{s+5/2} f(x,e),
 \label{R1}
\end{equation}
where, for whichever of the models described by the power-law density
profile,
\begin{equation}
 f(x,e)= 
 \frac{\left(1+e^2-x\right)\left(1-e^2\right)^{s+5/2}}
 {1+\frac{73}{24}e^2+\frac{37}{96}e^4}\;
 \sqrt{2+e^2-2x \over 1-x^2}\,.
 \label{f}
\end{equation}
For the standard thin disc model (i) we obtain
\begin{eqnarray}
 {\calR}_{\rm{}col/gw}^{\rm{}(i)} &\approx& 1.8 \times 10^{-2}M_8\mue^{-1}
 \left[ \frac{\alpha}{0.1} \right]^{-1}
 \nonumber \\ & & \times
 \left[ \frac{\Sigma_\ast}{\Sigma_{\sun}} \right]^{-1}
 \left[ \frac{M_\ast}{M_{\sun}} \right]^{-1}
 \left[ \frac{a}{r_{\rm g}} \right]^{4} f(x,e)\,,
 \label{eq:f_inner}
\end{eqnarray}
while for a gas pressure dominated disc (ii)
\begin{eqnarray}
 {\calR}_{\rm{}col/gw}^{\rm{}(ii)}
  &\approx& 1.5 \times 10^{2}M_8^{6/5}\mue^{3/5}
 \left[ \frac{\alpha}{0.1} \right]^{-4/5}
 \nonumber \\ & & \times
 \left[ \frac{\Sigma_\ast}{\Sigma_{\sun}} \right]^{-1}
 \left[ \frac{M_\ast}{M_{\sun}} \right]^{-1}
 \left[ \frac{a}{r_{\rm g}} \right]^{29/10} f(x,e)\,.
 \label{eq:f_middle}
\end{eqnarray}
For the model (iii) we find
\begin{eqnarray}
 {\calR}_{\rm{}col/gw}^{\rm{}(iii)} &\approx& 8.6 \times 10^{2}M_8^{6/5}\mue^{7/10}
 \left[\frac{\alpha}{0.1}\right]^{-4/5}
\nonumber \\ & & \times
 \left[\frac{\Sigma_{\ast}}{\Sigma_{\sun}}\right]^{-1}
 \left[\frac{M_\ast}{M_{\sun}}\right]^{-1}
 \left[\frac{a}{r_{\rm g}}\right]^{7/4} f(x,e)\,.
 \label{eq:f_out}
\end{eqnarray}
We show the functional form (\ref{f}) in Fig.~\ref{fgr2} where the factor
$f(x,e)$ determining the dependence of ${\calR}_{\rm{}col/gw}$ on
inclination and eccentricity is plotted for the case
(\ref{eq:f_middle}). In other words, $f(x,e)$ represents that part of
the drag ratio that is independent of the disc medium and the satellite
physical properties; only the two mentioned orbital parameters play a
role here. Typically, for a solar-type star it is only on eccentric
orbits that $f$ becomes small enough to bring ${\calR}_{\rm{}col/gw}$
below unity. The required eccentricity is rather high, and such a
satellite would be trapped or disrupted directly by the central hole.
Otherwise, ${\calR}_{\rm{}col/gw}\gg1$ for $a{\gta}r_{\rm{g}}$ and for
standard values of the disc parameters ($\alpha\lta1$, $\mue\approx1$).
This means that direct hydrodynamical interaction with the disc plays a
dominant role in the orbital evolution of satellites crossing the disc,
unless the medium is extremely rarefied (e.g., an {\sf{}ADAF}; Narayan
2000). Notice that the point $f(1,0)=0$ (i.e.\ a fully circularized
orbit inclined into the disc plane) is the exception, where the adopted
approximation of instantaneous collisions breaks down.

Analogous to ${\calR}_{\rm{}col/gw}$, one could explore the relative ratio
of the hydrodynamical versus gravitational radiation losses in other
regimes of the satellite-disc interaction. In this way, the relevant
formulae (those which apply in the course of orbiter evolution) are
\begin{eqnarray}
 {\calR}_{\rm{}dw/gw} &\!=\!& 2.3\times 10^{-12} M_8 
 \mue^{-3} \left[ \frac{\alpha}{0.1} \right]^{-1}
 \left[ \frac{a}{r_{\rm{g}}} \right]^8 \!, 
 \label{R2} \\
 {\cal R}_{\rm{}gap/gw} &\!=\!& 5.2\times 10^4 M_8 \mue^2 
 \left[ \frac{\alpha}{0.1} \right]^{-1}
 \left[ \frac{M_\ast}{M_{\sun}} \right]^{-1} 
 \left[ \frac{a}{r_{\rm{g}}} \right]^{1/2} \!\!,
\end{eqnarray}
for the inner, radiation pressure dominated disc (i), and
\begin{eqnarray}
 {\cal R}_{\rm{}dw/gw} &\!=\!& 1.0\times 10^{-4} M_8^{7/5} \mue^{1/5} 
 \left[ \frac{\alpha}{0.1} \right]^{-2/5}
 \left[ \frac{a}{r_{\rm{g}}} \right]^{19/5} \!, \\
 {\cal R}_{\rm{}gap/gw} &\!=\!& 6.3\, M_8^{4/5} \mue^{2/5} 
 \left[ \frac{\alpha}{0.1} \right]^{4/5}
 \left[ \frac{M_\ast}{M_{\sun}} \right]^{-1} 
 \left[ \frac{a}{r_{\rm{g}}} \right]^{13/5} 
 \label{R3}
\end{eqnarray}
for the middle region (ii). Here, the eccentricity-dependent factor was
omitted upon the finding that orbits are almost circular when inclined in the
plane of the disc. It is evident from eqs.\ (\ref{R1})--(\ref{R3}) that
gravitational radiation can have a visible impact only at small $a$, especially
when eq.~(\ref{R2}) applies (cp.\ also Fig.~\ref{fgr6} and related discussion 
below).
\begin{figure}[t]
\centering
\includegraphics[width=\mylength]{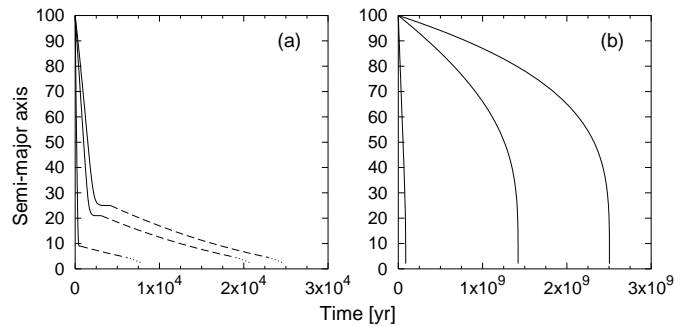}
\caption{Evolutionary tracks of the satellite are plotted in the 
plane semimajor axis $a$ (units of $r_{\rm{g}}$) versus time (years).
Two very different cases are shown.
(a)~Left: solar-type satellite ($\Sigma_{\ast}=\Sigma_{\sun}$). The
three curves correspond to different initial eccentricities (from top to
bottom), $e_0=0$, 0.4, and 0.8. Each curve is divided into three segments.
Solid line indicates the period of motion outside the disc (when
collisions occur with large orbital inclination); dashed line
corresponds to density-wave driven motion in the disc; dotted line is
the late stage with gap formation. (b)~Right: a compact satellite
($\Sigma_{\ast}=10^7\Sigma_{\sun}$). Only the first stage is resolved
here. In all cases, $a_0=100$, $x_0=0$. Gravitational radiation
contributes to orbital changes independent of inclination but its impact
is clearly important near the centre, where $a$ decreases rapidly.}
\label{fgr3}
\end{figure}

\subsection{Evolutionary tracks of the satellite}
Now we explore the evolutionary tracks of the satellite in the parameter
space of osculating elements. We start by considering two effects:
gravitational-wave losses in the approximation of eqs.\
(\ref{degw})--(\ref{dlgw}), and hydrodynamical drag acting on the
satellite according to eq.~(\ref{eq:velocity_change}) twice per
revolution. Dissipation operates with an efficiency depending on the type
of satellite, and it provides a mechanism for the separation of
different types of bodies in the phase space of a cluster.

Typical results of orbit integrations are presented in
Figure~\ref{fgr3}. Notice the big difference in time-scales relevant for
non-compact stars (left panel) when compared with compact ones (right
panel). In the former case, hydrodynamical drag is more pronounced. It
gradually changes the orbital plane, while gravitational radiation can be
safely neglected. The satellite sinks in the disc where the impulsive
approximation (\ref{dadtcoll}) loses its validity and it is substituted
by motion in the disc plane. Time-scales are generally longer in the
latter, compact satellite case, although gravitational wave emission
speeds up the evolution at very late stages ($a\lta50$). For
definiteness, we adopt the disc model (iii) and the condition of
$\tan{i_{\rm{}tr}}=h/r$ when the transition in the disc occurs. We
checked by modifying the value of $\tan{i_{\rm{}tr}}$ by a factor of $10$
that the qualitative picture of orbital evolution does not depend on
its exact choice and also that numerical results remain similar.

\subsection{Limitations of the adopted approximations}
For simplification, we ignored the effects of gradual change of the mass
of both the satellite and the central body. Furthermore, we assumed that
the interaction has no effect on the structure of the satellites. This
is a plausible assumption for stars with column densities much larger
than that of the disc, while it is inadequate for giants which must
quickly lose their atmospheres (Armitage et al.\ 1996). Also,
we did not consider various effects acting on the disc structure (e.g.\
torques imposed on it by the dense cluster of stars; Ostriker 1983).
Although all these effects will be important for a complete unified
treatment of accreting black holes in active galactic nuclei, they can
be neglected without losing the main physical effects influencing the
satellite motion in the present simplified scheme.

We note that the significance of direct orbiter's collisions with the
disc material is controlled by a characteristic time-scale
$\tau\,\propto\,\Sigma_{\ast}/\Sigma_{\rd}(r)$, which we expressed in
terms of the orbital decay $\dot{a}$. Let us recall that dimensionless
parameter characterizing compactness of the orbiter can be introduced in
different ways. While $\Sigma_{\ast}/\Sigma_{\rd}$ stands directly in
the description of star-disc collisions, the usual factor
$\varepsilon=GM_{\ast}R_{\ast}^{-1}c^{-2}$ determines the importance of
general relativity effects near the surface of a compact body:
$\varepsilon\approx10^{-6}\left(\Sigma_{\ast}/\Sigma_{\sun}\right)
\left(R_{\ast}/R_{\sun}\right)$. We treat motion in the Newtonian regime
and we only take the possibility of the satellite capture into account
by removing the orbiter from a sample if its trajectory plunges too
close to the central mass, below a marginally stable orbit. Another
dimensionless quantity has also been designated as the compactness
parameter when describing accretion onto compact objects,
$\tilde{\varepsilon}=L\sigma_{_{\rm{}T}}r^{-1}m_{\rm{}e}^{-1}c^{-3}$. 
It considers the effect of radiation luminosity $L$ acting through a
cross-section $\sigma_{_{\rm{}T}}$ in the medium, however, we can safely
ignore radiation pressure on macroscopic satellites hereafter.

The gradual and monotonic decrease of eccentricity is overlaid with
short-term oscillations if the disc mass is non-negligible
(Vokrouhlick\'y \& Karas 1998). Also the satellites' inclination
converges to a somewhat different distribution (instead of a strictly
flattened disc-type system) when two-body gravitational relaxation is
taken into account (\v{S}ubr 2001; Vilkoviskiy 2001 -- preprint). On the
other hand, complementary to the scenario of the satellites grinding
into the disc is the picture of enhanced star formation in the disc
plane, in which case the stars are born with zero inclination (Collin \&
Zahn 1999). But these effects, as well as evaporation processes
operating in the cluster, as suggested by various Fokker-Planck
simulations (e.g., Kim, Morris \& Lee 1999),  remain beyond the scope of
the present paper.

We could see that different mechanisms (of which we considered
particular examples) affect the orbital motion rather selectively,
depending on the orbiter's size and mass. One thus expects separation of
different objects in the cluster phase space. In order to verify this
expectation we examine in the following paragraph a simple scheme which
captures gradual changes in the structure of the cluster. Such a
discussion is required: indeed, in the absence of sufficient
resolution which would enable tracking of individual stellar paths in
nuclei of other galaxies, one needs to inspect the overall influence on
the members of the cluster, namely, the change of the radial
distribution of the satellites in terms of their number fraction and
average inclination.

Let us note that the accretion flow is supposed to remain undisturbed
by the presence of the embedded cluster. This assumption gives an upper
limit on the total number of stars inside the radius $r_{\rm{d}}$
($\approx10^4r_{\rm{}g}$), and on the fraction of those dragged into the
disc plane in this region. A simple smooth disc can be destroyed,
especially in the process of gap formation (the case of sufficiently
large $M_{\ast}$ and small $h$); the models (i) and (v) are particularly
susceptible to the occurence of multiple gaps. Very roughly, if $\approx2\%$ of
the total number of $N=10^4$ satellites get aligned with the disc at late times
(a result of our computations for the model (iii)), then their Roche lobes 
might cover an area of the order $\approx0.01Nr_{\rm{d}}r_{_{\rm{}L}}$.
This is just comparable with the total disc surface for 
$M_{\ast}\approx1M_{\sun}$, however, recall that
there are more conditions for the gap formation depending on the disc
model. Also, effects of two-body relaxation and of satellite scattering
in the gravitational potential of the disc are expected to decrease
the fraction of aligned bodies when these effects are taken into account.

\section{The cluster evolution}
\label{evolution}
We follow the cluster evolution starting from a clean stationary state
(no dissipative perturbation), which corresponds to a spheroidal stellar
system gravitationally coupled with the central black hole (Bahcall \&
Wolf 1976). Hence, number density $n_0(a)$ of the satellites is assumed 
proportional to $a^{1/4}$, while the initial distribution in eccentricities 
and inclinations conforms with the condition of a randomly generated sample. 
Common parameters of
the computations are the mass $M=10^8M_{\sun}$ of the central body, and
the orbiters' column density $\Sigma_{\ast}=\Sigma_{\sun}$. Now, the
perturbation effects are switched on, depending on the prescribed disc
type and accretion rate (typically $\mue=1$, but we also examined other
situations and less conventional values of the above parameters; cf.\
\v{S}ubr 2001).

As a consequence of the dissipative action, the system departs from the
initial configuration. We concentrate our main attention on the range of
radii $r\lta10^3r_{\rm{g}}$, well below the outer edge of the disc
$r_{\rm{}d}\approx10^4r_{\rm{g}}$ (where $\Sigma_{\rd}$ becomes
negligible). We find a new quasi-stationary, modified cluster
distribution developing in that region. On the outer boundary, fresh
satellites could be inserted in the system from an external reservoir in
order to maintain the steady flow towards the center, but in this work
we alternatively consider a closed system with a large but finite number
of its members, $N_0$, which are not replenished. The actual value of
$N_0$ is not very important in the present work because we neglect
gravitational interaction among the satellites themselves ($N_0$ will
play a role when two-body collisions are taken into account).

\subsection{A role of the disc model}
Several different models of the disc were adopted. We specify them in
terms of their corresponding density profiles, $\Sigma_{\rd}(r)$, and
vertical thicknesses, $h(r)$:

\vspace{3pt}\small

\hspace*{\fill}\parbox{0.9\mylength}{(i)~Radiation pressure dominated,
geometrically thin disc (Shakura-Syunaev $\alpha$-viscosity model; Frank
et al.\ 1992)}\vspace{1pt}

\hspace*{\fill}\parbox{0.9\mylength}{(ii)~Gas pressure dominated
standard disc with opacity due to electron scattering.}\vspace{1pt}

\hspace*{\fill}\parbox{0.9\mylength}{(iii)~The same as in (ii) but
with opacity dominated by free-free scattering.}\vspace{1pt}

\hspace*{\fill}\parbox{0.9\mylength}{(iv)~Gravitationally unstable
discs in the region of a self-gravitating disc beyond the Toomre radius;
see Sec.~(2.3) of Collin \& Hur\'{e} (1999) for analysis of the
marginally unstable ($\zeta=5$) self-gravitating disc with solar
metallicity and large optical thickness.}\vspace{1pt}

\hspace*{\fill}\parbox{0.9\mylength}{(v)~The same as in (iv) but for
optically thin medium with zero metallicity.}\vspace{3pt}

\normalsize\noindent{}The choice of different models is dictated by
their applicability for the description of accretion regimes relevant
under different conditions, namely, the cases (iv)--(v) are appropriate
for black-hole discs at large ($\gta10^3r_{\rm{}g}$) radii. One could
consider other models relevant in various situations but the adopted
examples cover qualitatively different evolutions which are frequently
encountered. In particular, the above models comprise a range of
$\Sigma_{\rd}/\Sigma_{\ast}$ dependencies (dimensionless factor
determining the evolutionary time-scales) and different
$(h/r)(\Sigma_{\rd}/\Sigma_{\ast})=\epsilon/\tau$ whose radial
dependence determines area in the disc where the gap is formed.

First, we carried out simulations while sticking to one of these models
in the whole range of radii from the centre up to the outer edge, even
if it extends somewhat out of the zone of validity of the particular
model. This helps us to avoid additional complexities which could stem
from a complicated prescription for the disc properties. For example,
one expects a different rate of radial drift across boundaries where the
disc properties are switched, but we refrain from this complication for
the moment. This effect will be observed later on: to achieve a more
realistic description, a further step will be carried out by joining
different regions of the disc in a unified model. In that case, the
radiation pressure supported standard disc might be limited to the
innermost part of the system while gravitationally unstable part takes
over at distances of the order of a few hundreds $r_{\rm{g}}$.

Parameters defining radial migration in the models (i)--(v) were 
given in Table~\ref{tab1}. Here we
only remark that the factor $B$ determines the pace in absolute time units
with which the evolution proceeds, while power-law indices $q_i$
influence the form of the satellites number density distribution $n(a)$.

\begin{figure}[t]
\centering
\includegraphics[width=\mylength]{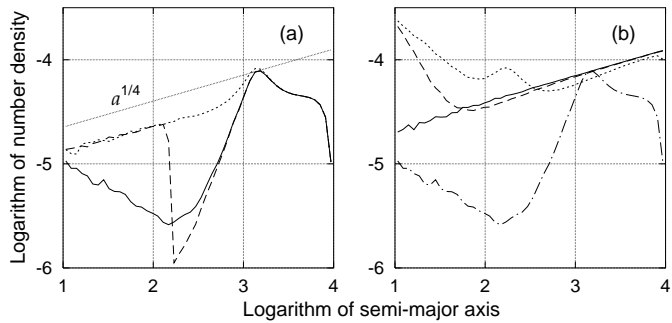}
\caption{Number densities $n_t(a)$ of the cluster. Left
panel: quasi-stationary states for $M_{\ast}=1M_{\sun}$ (solid line),
$3M_{\sun}$ (dashed), and $10M_{\sun}$ (dotted). Right panel:
snapshots of $n_t(a)$ are shown for $M_{\ast}=1M_{\sun}$ at $t=10^6$
(solid), $10^8$ (dashed), $10^{10}$ (dotted), and $10^{12}$ 
(dash-dotted). In all cases, $\Sigma_\ast=\Sigma_{\sun}$ (solar-type
compactness of the satellite). On the vertical axis, units are arbitrary
and scaled with respect to the initial profile $n_0(a)\,\propto\,a^{1/4}$. 
Units of $GM/c^2$ are used for $a$ on the horizontal axis.}
\label{fgr4new}
\end{figure}

\subsection{Modified cluster distributions}
\label{sec:modified}
The main results are shown in Figure~\ref{fgr4new}, in which the disc
model (iii) is adopted. This will serve as a representative test case
(more examples are deferred to Appendix). It is particularly interesting
to observe transition regions where the mode of the mutual
satellite-disc interaction is changed. At the transition, the efficiency
of radial transfer of the satellites is changed, and this effect
produces sudden changes (either dips or concentrations) in $n_t(a)$
profiles. If the average inclination of the satellites is small (the
case of a flattened configuration), a ring is created in the place where
the orbital decay is stalled.

The computations were launched from the initial distribution
$n_0(a)\,\propto\,a^{1/4}$. In the left panel, Fig.~\ref{fgr4new}a, new
$n_t(a)$ profiles are shown at a sufficiently late time, $t=10^{12}$, when
a quasi-stationary state has been established.\footnote{Time is given in
units $GM=c=1$, total number of satellites is $N=\int{}n_t(a)\,\rd{a}$.
Conversion of time $t$ to physical units is achieved by
$t_{\rm{}phys}\mbox{[yr]}=1.6\times10^{-5}M_8t$, while for distances
$a_{\rm{}phys}\mbox{[cm]}=1.5\times10^{13}M_8a$.}
\setcounter{mynumber}{\value{footnote}} Three cases are plotted with
increasing satellite masses: $M_{\ast}=1M_{\sun}$, $3M_{\sun}$, and
$10M_{\sun}$. After being inclined to the disc, low mass satellites
proceed via density waves (e.g., $1M_{\ast}$ case with
$n_t(a)\,\propto\,a^{-1/2}$ at small $a$). On the other hand, very
massive satellites develop a gap already very far from the centre and
they proceed in this mode the whole way down ($10M_{\ast}$ case with
$n_t(a)\,\propto\,a^{1/4}$, which is incidentally parallel with the
initial $n_0$ distribution).\footnote{The values of $n_t(a)$ slopes are
further explained in Sec.~\ref{sec:qs}. We recall that these slopes refer
to the motion in the disc. The overall $n_t(a)$ profiles can be thus
different in the region where the whole sample is dominated by
those satellites moving along inclined trajectories outside the disc.}
Intermediate satellites show an evident dip in $n_t(a)$: its formation
is connected with the gap in the disc which arises at sufficiently small
$a$. Exact location of the dip depends on the disc model too; for
$3M_{\sun}$ satellites it occurs at $a\lta150$.

The right panel, Fig.~\ref{fgr4new}b, is complementary to the left one,
showing temporal evolution of $n_t(a)$ for one solar mass satellites.
One notices two areas with different $n_t(a)$ slopes in the graph: at
small $a$ the overall distribution is eventually dominated by satellites
residing in the disc (corresponding to negative slope), while a whole
mixture of satellite inclinations persists farther out (a positive
slope). Transition between the two regions occurs at $a\approx10^2$ (its
location moves gradually towards larger $a$ as time proceeds; see the
figure). A terminal dip is seen at large $a$ (on the right of the
graphs); it is caused by depletion of the sample in the simulation.

A realistic situation must involve further changes of the $n_t(a)$ slope
caused by varying the accretion mode (see Appendix and Fig.~\ref{fgr7} for
details). In this respect, the form of quasi-stationary distributions is
of particular interest (the persisting quasi-stationary profile of
$n_t(a)$ is given more detailed discussion in the next section,
Sec.~\ref{sec:qs}). We remark that the long-term evolution of the orbits
is influenced by collisions with the disc especially if the satellite's
cross-sectional area is large, e.g.\ a giant star or a solar-size body,
but quite the opposite conclusion can be drawn for compact objects such as
neutron stars and stellar-mass black holes for which the cross-section
is much smaller, diminishing their collisional interaction with the
disc. The whole cluster is modified by this kind of dissipative process
in a selective manner, depending on $M_{\ast}$.

\begin{figure}[t]
\centering
\includegraphics[width=\mylength]{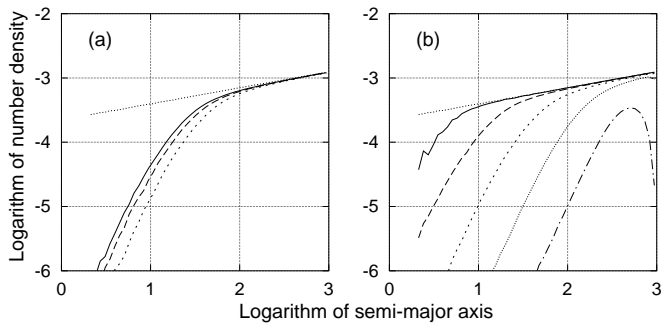}
\caption{Left panel: the shape of the distribution $n_t(a)$ of a
cluster containing compact orbiters of the masses $M_{\ast}=1.5M_{\sun}$
(solid line), $3M_{\sun}$ (dashed) and $10M_{\sun}$ (dotted) at
$t=10^{12}$. Right panel: subsequent phases of evolution for a cluster 
of neutron stars with $M_{\ast}=1.5M_{\sun}$. The power-law is 
represented by a line in the log-log plane of $n_t$ (normalized by $N_0$, 
the initial number of cluster members) vs.\ $a$ (units of $r_{\rm{}g}$). 
Again, a thin dotted line stands for the initial power-law distribution, 
$n_0$. Other curves correspond, from top to bottom, to $t=10^9\!$, $10^{11}\!$, 
$10^{13}\!$, $10^{15}\!$, and $10^{17}\!$.}
\label{fgr5}
\end{figure}

In Figure~\ref{fgr5}a we present the resulting form of the distribution
$n_{t=10^{12}}(a)$ for different masses of compact bodies. On the
other hand, panel \ref{fgr5}b concerns a cluster consisting of one
solar-mass satellites, $M_{\ast}=1M_{\sun}$, shown at different times
(several consecutive moments are examined with a logarithmic spacing in
$t$). Now, there is no need to consider separately that part of the
distribution of trajectories inclined into the disc; indeed, no compact
bodies reach the disc plane, even at late phases of the evolution. This
implies that the orbits evolve predominantly by gravitational radiation
and the resulting graphs come out almost identical for all the disc
models under consideration. We are thus led to restrict ourselves to
non-compact satellites which suffer relatively intense drag from the
disc.

\subsection{A quasi-stationary state}
\label{sec:qs}
Time-scales of the perturbed cluster evolution depend on the disc model
and, as shown above, a number of other factors characterizing the cluster
members -- their typical mass, compactness and efficiency of the
dissipative interaction with the disc medium. With solar-type (and less
compact) satellites, the disc plays a dominant role in determining the
form of $n_t(a)$. The starting canonical distribution
$n_0(a)\,\propto\,a^{1/4}$ (corresponding to $\rho\,\propto\,r^{-7/4}$
in the notation of Bahcall \& Wolf 1976) changes to $n_t(a)$, where it
stands almost frozen in a quasi-stationary state for a 
prolonged interval of time. This is true also in the case 
of initial profiles reasonably different from the
canonical one. In particular, we verified that $n_0(a)={\rm{}const}$
converges to identical slopes as those shown in Fig.~\ref{fgr7} at
$t\approx10^{12}$ (cf.\ the bottom row of plots in the Appendix).

Clear-cut results emerge provided the distribution is influenced by a
single mode of star-disc interaction. The evolution time-scales are
shorter in the region where substantial fraction of the satellites are
already inclined into the disc plane, so that $n_t(a)$ forgets about the
exact initial distribution much earlier, at about $t\gta10^8$--$10^9$.
In that case the final slope $q_{\mathrm{f}}$ ranges from the initial
value up to the limiting value $q_4$ of the corresponding model and
the mode of migration (the
values are listed in Table~\ref{tab1}). The actual final value of
$q_{\mathrm{f}}$ reaches $q_4$ if two conditions are satisfied in the
region where the power-law distribution is established: first, the flow of
the satellites in this region continues in the disc plane (all bodies
are inclined to the disc); second, the satellites inflow is faster than
the changes of their number density at the outer boundary. These
conditions are satisfied in the late stages until the sample is
exhausted. In other words, the modified cluster distribution reflects
the adopted disc model while it is rather insensitive to the initial
form of $n_0(a)$.
\begin{figure}[t]
\centering
\includegraphics[width=\mylength]{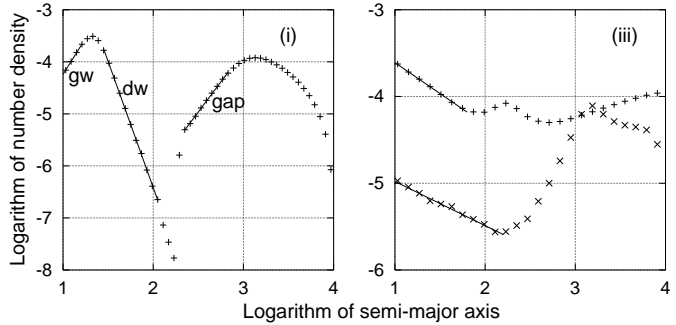}
\caption{Left panel: a quasi-stationary $n_t(a)$ profile is plotted (by
crosses) for the model (i), $t=10^{13}$. A power-law approximation,
$n_t(a)\,\propto\,a^{q_{\mathrm{f}}}$, is constructed in appropriate
ranges of radius (solid lines). Suppression of the profile occurs at small
$a$ in consequence of quite efficient gravitational radiation losses
near the center. Right panel: the slope evolves from the initial value,
$q=1/4$, until the quasi-stationary state of
$q\rightarrow{q_{\mathrm{f}}}$ is reached. Two snapshots are shown for
the model (iii): $t=10^{10}$ (upper profile), and $t=10^{12}$ (lower
profile). See the text for corresponding values of $q_{\mathrm{f}}$.}
\label{fgr6}
\end{figure}

This behaviour is illustrated in Figure~\ref{fgr6} in terms of a
power-law approximation to the computed distributions. Given the disc
model, different power-law relations may be established in separate
regions depending on the mode of star-disc interaction applicable at
that distance. In the case (i) and for solar-type satellites, we find
that the slope $q_{\mathrm{f}}=-4.9$ arises in the region
$30{\lta}a{\lta}120$ (the evolution is driven by density waves,
accumulating the satellites in an evident peak), while
$q_{\mathrm{f}}=2.3$ suits for $220{\lta}a{\lta}680$ (i.e.\ beyond the
transition point where the star-disc interaction is determined by the
gap; here, as an example, $q_{\mathrm{f}}=2.3$ is to be compared with
the corresponding $q_4=\slf{5}{2}$ in the sixth row in Table~\ref{tab1}).
Furthermore, the slope is influenced by gravitational radiation which
dominates the evolution on very small radii: $q_{\mathrm{f}}=2.8$ for
$r\leq20$. The three regions are distinguished in the graph: 
{\sf{}gw} (dominated by gravitational waves), {\sf{}dw} (density waves), 
{\sf{}gap} (gap formation). The large span of values acquired by 
$q_{\mathrm{f}}$ reflects the presence of ripples in $n_t(a)$ which develop 
in certain areas of the disc. 

The right panel of Fig.~\ref{fgr6} shows how the $n_t(a)$ profile is
gradually adjusted as time proceeds. Here we plot the case (iii) with
$M_{\ast}\lta1M_{\sun}$ for illustration, and we find $q=-0.71$
($t=10^{10}$, $10\lta{r}\lta60$), and $q=-0.51$ ($t=10^{12}$,
$10\lta{r}\lta160$), respectively. On the vertical axis, one can read
the fraction of satellites with semi-major axes falling in interval
$\langle{a,a+{\rm{}d}a}\rangle$. One observes different slopes for 
more massive satellites, $M_{\ast}\gta3M_{\sun}$, because 
of gap formation (the slope reaches $q\dot{=}0.25$); cp.\ with 
Fig.~\ref{fgr4new} to see the effect of $M_{\ast}$. On the other 
hand, the situation gets simpler with the other models where such a
transition does not occur, and a single power-law relation prevails at
late stages for all $M_{\ast}\lta10M_{\sun}$. The emerging slopes are: (ii)~$q_{\mathrm{f}}=-0.79$, and (iv)~$0.30$, for which we found that the 
quasi-stationary state is established at $r\lta200$, $t=10^{12}$. The case (v)
develops more slowly and reaches the expected value $q_{\mathrm{f}}=-0.67$
somewhat later,\footnote{The expected final value of $q_{\mathrm{f}}$ 
for the case (v) coincides with $q_4=-\slf{2}{3}$, as can be seen in the 
corresponding section ({\sf{gap}}) of Tab.~\ref{tab1}.} at time 
$t=10^{13}$.

\section{Conclusions}
Energy and angular-momentum losses of a satellite
colliding with an accretion disc were examined. We explored how the relative
importance of several secular effects depends on the parameters of the
model, namely, the osculating elements of the satellite trajectories,
surface density of the disc, and the mass and compactness of the
orbiter bodies. We verified within the thin-disc analysis that
hydrodynamical drag is, typically, more important for the long-term
orbital evolution than gravitational radiation losses. Decay of the
orbit due to interaction with the gaseous environment brings the orbiter
gradually towards the center. This drag acts at different regimes
depending on parameters of the orbiters and of the disc. The magnitude of
the drag determines the rate of orbital decay and it influences the
resulting structure of the cluster, especially the characteristic slope
of $n_t(a)$ density distribution.

Gravitational radiation losses dominate over the hydrodynamical
dissipation if the disc has low density and/or the orbiter is of very
high compactness (a neutron star or a black hole). The two influences
are thus complementary and they can operate simultaneously at different
regions of a given source, assuming the thin-disc scheme is valid
farther away from the center, whereas advection-dominated flow takes over
at distances of the order of $\approx10^2r_{\rm{}g}$. One could thus expect
the final stages of the orbital decay to be governed by emission of
gravitational waves and by tidal interaction, while the initial transport
from the outer cluster is ensured by other causes, possibly the
hydrodynamical collisions together with dynamical friction. Our results
are still only indicative in this part because several effects were
ignored which must be taken into account in realistic models, namely,
distribution of the satellite orbital parameters may be more
complicated if two-body collisions between the orbiters are considered.

Another open question is the impact of star-disc interactions on the
mass function of the satellites. We observed that $M_{\ast}$, 
$\Sigma_{\ast}$ and $\Sigma_{\rd}$ are the crucial parameters 
determining the cluster evolution, but the parameter range spans an
enormous interval for different types of objects. Compact bodies have
$\Sigma_{\ast}\approx10^9\Sigma_{\sun}=1.3\times10^{20}\,{\rm{}g\,cm^{-2}}$
and hence they are affected less by collisions with the disc than
solar-type stars, which are aligned with the disc 
plane more rapidly. Consequently, the initial mass function is
modified towards a higher abundance of compact stars residing in inclined
orbits, and vice versa. However, in the relatively dense environment of
the disc, the stars accrete at an enhanced rate, they will soon gain
sufficient mass and eventually collapse, producing additional compact
bodies. A detailed discussion of their subsequent evolution under the
influence of the disc environment remains beyond the scope of the
present discussion. We only remark that solar-mass satellites can
substantially multiply their masses during $10^6$ years (Collin \& Zahn
1999), assuming that their own radiation and the effect of gaps do not 
halt further accretion. This is almost two orders of magnitude shorter 
than the expected quasar life-time (Haehnelt \& Rees 1993) and hence the
effects of star-disc interaction should not be neglected.

As a concluding remark, let us recall {\em{}de nouveau\/} that the final 
stages of an orbiter located near the center are relevant for 
gravitational wave experiments. The radiation losses govern the orbital
evolution if the satellite is compact enough; on close orbits 
($e\approx0.9$, $a\approx10^2r_{\rm{g}}$) the influence of 
gravitational radiation is comparable to the effects of star-disc 
collisions, even if the medium is relatively dense, $\Sigma_{\rd}\approx
10^5\,{\rm{}g\,cm^{-2}}$.

\begin{acknowledgements}
The authors acknowledge discussions with Suzy Collin about gravitationally
unstable regions of accretion discs, and useful comments and suggestions
by the referee, who helped us to improve clarity of the text. Support 
from the grants GAUK 188/2001, GACR 205/00/1685, and 202/99/0261 is also 
gratefully acknowledged.
\end{acknowledgements}

\appendix
\section{Details on modified clusters}
\begin{figure*}
\centering
\includegraphics[width=\textwidth]{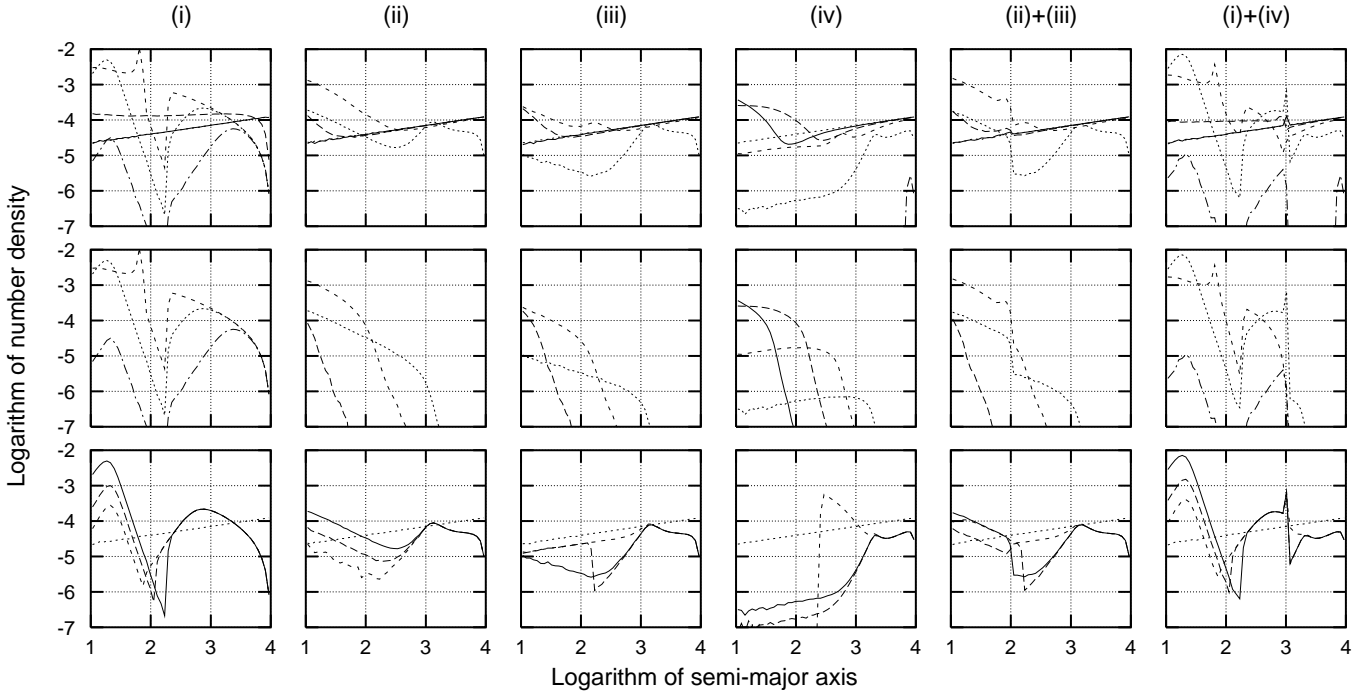}
\caption{A modified distribution $n_t(a)$ of satellites in the sample.
Columns (i)--(iv) show the modified distributions which are developed
following the interaction with the corresponding disc models. (The
models are listed in Sec.~\ref{evolution}.) The last two columns reflect
the structure of composite models consisting of different regions, as
indicated on top. The evolution of the composite models can be compared
with the case of individual simple models from which they are
constructed. In each graph, line types of the curves reflect different
views of $n_t(a)$ profiles (see the text). First, snapshots are shown 
with varying time
(top row; $M_{\ast}=1M_{\sun}$); second, the fraction of satellites aligned
with the disc is extracted from the whole sample (middle row); third,
the quasi-stationary states are plotted for different $M_{\ast}$ (bottom
row). See the Appendix for exact interpretation of different
rows of graphs.}
\label{fgr7}
\end{figure*}
We can expand the results of Sec.~\ref{sec:modified} in slightly more
detail by examining how the modified cluster distribution depends on the
adopted accretion regime and how the quasi-stationary state is
eventually established. There is a qualitative difference from the
simpler case discussed in the main text, because varying surface density
profiles of the disc models (i)--(v) introduce potentially observable
concentrations of satellites at certain radii. Also, it is instructive
to recover in the graphs different influences resulting from orbital
decay, discussed for individual satellites in previous paragraphs
(Sec.~\ref{sec:individual}).

In Figure \ref{fgr7}, the first four columns correspond to simple models
adopted from our list. In the remaining two columns, the disc consists
of a conjunction of two different models, joined at a given radius: The
model denoted (ii)+(iii) is formed by the inner (ii) and the outer (iii) part 
of the gas pressure dominated disc with the transition at $100r_{\rm{}g}$.
Similarly, the model denoted (i)+(iv) is a conjunction of the radiation
pressure dominated disc spreading up to $10^3r_{\rm{}g}$ with the outer,
self-gravitating disc (Collin \& Hur\'{e} 1999). The model parameters 
determine the final
form of plotted curves in their mutual interplay together with the mass
$M_{\ast}$ of the cluster members, which remains the
characteristic influencing the efficiency of individual contributions
to the cluster evolution.

Given $a$, the values of $n_t(a)$ are distinguished from each other if
a substantial fraction of the satellites are inclined in the plane of the
disc without opening a gap in it. In such a situation the migration is
faster for more massive stars whose number density is then lowered with
respect to low-mass stars. Radial distribution of the satellites is
influenced by the mode of their migration in the disc, and as a
consequence the satellites are transported at different rates
towards the centre. As a result, a wiggle occurs in the $n_t(a)$ curve. The
particular value of $a$ where the satellites are accumulated at an
increased rate, forming a ring-type structure, depends on the details of
the model. However, the trend to accumulation at some distance is seen
in various other situations and it leads to a bias in the mass function
of the cluster.

For example, when the model (ii) is considered, the stars proceed under
the regime of density wave excitation in the disc plane. The
critical point (below which the gap would have been opened) lies at too
small distances, $\lta2r_{\rm{}g}$, even for the largest considered mass
$M_\ast=10M_{\sun}$. Another situation develops in the case (iii) 
where the gap is opened below $a=5$, $150$ and $7\times10^3$ for our 
three adopted values of satellite masses. This implies different shapes
of $n_t(a)$ for $M_\ast=M_{\sun}$ satellites with respect to more massive 
ones. Furthermore, in the marginally 
unstable disc, i.e.\ the case (iv), the point above which the gap could 
be opened is beyond the outer boundary of the considered region and, 
therefore, the distributions $n_t(a)$ are sensitive to the mass of the 
satellites (most of them are inclined in the disc at this stage).

In order to maintain clarity of the graphs we omit the curves
corresponding to the model (v). This case is similar to (ii)
except for the fact that the gap opens for all three considered stellar
masses, which makes the case (v) distributions very similar to each
other, quite independent of $M_{\ast}$. 

We could see that simple power-law profiles $\Sigma_{\rd}(r)$ lead to
different distributions with respect to more realistic models, which are
governed by different processes dominating at the corresponding radius.
In discs that are a compound of several parts, the surface density profile
contains several transitions between different regions, and hence it is
more complicated than a simple power-law. The resulting effects on the
boundaries resemble the transitions which we observed previously with
simple discs when the mode of radial migration is changed. A local peak
of $n_t(a)$ occurs at radii where the satellites enter the disc region
with slower radial motion, which is the situation of the composite model
(ii)+(iii), or quite on the contrary a dip develops in $n_t(a)$ as it
can be seen in the case (i)+(iv).

We observed that a quasi-stationary distribution $n_t(a)$, different
from the initial one, was established below $\lta10^3r_{\rm{}g}$ for
rather extended periods of time. However, a decline of $n$ eventually
arrives from large $a$ at final stages when the supply of satellites is
exhausted. The sample is eventually depleted. This can be also seen in
Fig.~\ref{fgr7}, the first (top) row, where time evolution is presented
for one solar-mass satellites. Dimensionless times are $t=10^6$ (solid
line), $10^8$ (long-dashed), $10^{10}$ (short-dashed), $10^{12}$
(dotted), and $10^{14}$ (dash-dotted). See footnote~\themynumber\ for
time units. The last curve is not visible in all plots because of a very
large $t$; it can be noticed e.g.\ in the case (iv) where the evolution is
slow and a fraction of satellites persists above $a\gta2\times10^3$.

Complementary to the above-described panels is the second row of graphs
where only a part of $n_t(a)$ distribution is plotted, corresponding to
the satellites aligned with the disc. Line styles denote time in the
same manner as in the first row. Naturally, the population of aligned
bodies is less frequent, and hence some curves are missing in
the second row with respect to their matching curves in the first row,
if no satellites are brought to zero inclination. 

Finally, the third (bottom) row shows $n_t(a)$ profiles taken at
$t=10^{12}$. At this moment a quasi-stationary state is already
established in the form depending on the mode of satellite-disc
interaction. The profiles are plotted for three different values of the
satellite mass, corresponding to $M_{\ast}=1M_{\sun}$ (thick solid
lines), $M_\ast=3M_{\sun}$ (long-dashed lines), and $M_\ast=10M_{\sun}$
(short-dashed lines). For example, one can check in column (ii) that the
structure of the plot is ruled by the density wave regime and acquires
the corresponding slope. (The initial slope 1/4 is also indicated with 
a light dotted line.) The curves
clearly reflect the fact that the orbital decay (\ref{dadtdw}) depends
on $M_{\ast}$.


\begin{thebibliography}{}
\bibitem{cit100}Armitage P.\,J., Zurek W.\,H., Davies M.\,B.\ 1996,
 ApJ 470, 237; astro-ph/9605137
\bibitem{cit110}Artymowicz P.\ 1994, ApJ 423, 581
\bibitem{cit120}Artymowicz P., Lin D.\,N.\,C., Wampler E.\,J.\ 1993, ApJ 409, 592
\bibitem{cit130}Bahcall J.\,N., Wolf R.\,A.\ 1976, ApJ 209, 214
\bibitem{cit140}Bekki K.\ 2000, ApJ 540, L79
\bibitem{cit150}Chakrabarti S.\,K.\ 1993, ApJ 411, 610
\bibitem{cit160}Collin S., Hur\'{e} J.-M.\ 1999, A\&A 341, 385
\bibitem{cit170}Collin S., Zahn J.-P.\ 1999, A\&A 344, 433
\bibitem{cit180}Colpi M., Mayer L., Governato F.\ 1999, ApJ 525,
 720; astro-ph/9907088
\bibitem{cit190}Dhurandhar S.\,V., Vecchio~A.\ 2001, Phys.\ Rev.~D 63,
 122001; gr-qc/0011085
\bibitem{cit200}Ferrarese L., Merrit D.\ 2000, ApJ 539, L9
\bibitem{cit210}Frank J., King A.\,R., Raine D.\,J.\ 1992, {\it{}Accretion Power in
 Astrophysics} (Cambridge University Press, Cambridge)
\bibitem{cit220}Haehnelt M.\,G., Rees M.\,J.\ 1993, MNRAS 263, 168
\bibitem{cit230}Hughes S.\,A.\ 2001, Phys.\ Rev.~D, in press;~gr-qc/0104041
\bibitem{cit240}Ivanov P.\,B., Papaloizou J.\,C.\,B., Polnarev A.\,G.\ 1999, 
 MNRAS 307, 79
\bibitem{cit250}Kato S., Fukue J., Mineshige S.\ 1998, {\it{}Black-Hole Accretion
 Disks} (Kyoto University Press, Kyoto)
\bibitem{cit260}Kim S.\,S., Morris M., Lee H.\,M.\ 1999, ApJ 525, 228
\bibitem{cit270}King A.\,R., Done C.\ 1993, MNRAS 264, 388
\bibitem{cit280}Lin D.\,N.\,C., Papaloizou J.\,C.\,B.\ 1986, ApJ 309, 846
\bibitem{cit290}Nakamura~T., Oohara~K., Kojima~Y.\ 1987, Progr.\ Theor.\
 Phys.\ Suppl.\ 90,~1
\bibitem{cit300}Narayan R.\ 2000, ApJ 536, 663; astro-ph/9907328
\bibitem{cit310}Ostriker E.\,C.\ 1999, ApJ 513 252; astro-ph/9810324
\bibitem{cit320}Ostriker J.\,P.\ 1983, ApJ 273, 99
\bibitem{cit330}Peters P.\,C., Mathews J.\ 1963, Phys.\ Rev.\ 131, 435
\bibitem{cit340}Pineault S., Landry S.\ 1994, MNRAS 267, 557
\bibitem{cit350}Rauch K.\,P.\ 1995, MNRAS 275, 628
\bibitem{cit360}Rauch K.\,P.\ 1999, ApJ 514, 725
\bibitem{cit370}Rees M.\,J.\ 1998, in {\it Black Holes and Relativistic Stars},
 ed.\ R.\,M.\ Wald (University of Chicago Press, Chicago), p.~79
\bibitem{cit380}Shlosman I., Begelman M.\,C.\ 1989, ApJ 341, 685
\bibitem{cit390}Shlosman I., Noguchi M.\ 1993, AJ 414, 474
\bibitem{cit400}Sridhar S., Touma J.\ 1999, MNRAS 303, 483; astro-ph/9811304
\bibitem{cit410}Syer D., Clarke C.\,J., Rees M.\,J.\ 1991, MNRAS 250, 505
\bibitem{cit420}\v{S}ubr L.\ 2001, Thesis (Charles University Prague), in
 preparation
\bibitem{cit430}\v{S}ubr L., Karas V.\ 1999, A\&A 352, 452; astro-ph/991040
\bibitem{cit440}Takeuchi T., Miyama S.\,M., Lin D.\,N.\,C.\ 1996, ApJ 460, 832
\bibitem{cit450}Tremaine S.\ 1995, AJ 110, 628
\bibitem{cit460}van der Marel R.\,P., van den Bosch F.\,C.\ 1998, AJ 116, 2220
\bibitem{cit470}Vokrouhlick\'y D., Karas V.\ 1993, MNRAS 265, 365; astro-ph/9305035
\bibitem{cit480}Vokrouhlick\'y D., Karas V.\ 1998, MNRAS 298, 53; astro-ph/980501
\bibitem{cit490}Ward W.\ 1986, Icarus 67, 164
\bibitem{cit500}Ward W.\ 1997, Icarus 126, 261
\bibitem{cit510}Zurek W.\,H., Siemiginowska A., Colgate S.\,A.\ 1994, ApJ 434, 46
\bibitem{cit520}Zwart S.\,F.\,P., Hut P., Verbunt F.\ 1997, A\&A 328, 130
\end{thebibliography}
\end{document}